\documentclass{aastex}
\usepackage{amsmath}
\usepackage{natbib}

\def\edth{\;\raise1.0pt\hbox{$'$}\hskip-6pt\partial\;}
\def\baredth{\;\overline{\raise1.0pt\hbox{$'$}\hskip-6pt
\partial}\;}
\def\gsim{~\rlap{$>$}{\lower 1.0ex\hbox{$\sim$}}}
\def\lsim{~\rlap{$<$}{\lower 1.0ex\hbox{$\sim$}}}

\def\MNRAS{Mon. Not. R. Astron. Soc.}
\def\ApJ{ ApJ}

\def\be{\begin{equation}}
\def\ee{\end{equation}}
\def\bea{\begin{eqnarray}}
\def\eea{\end{eqnarray}}

\def\lsim{~\rlap{$<$}{\lower 1.0ex\hbox{$\sim$}}}
\def\bigint{\begingroup \displaystyle
 \edef\@bi@fam{\fam\the\fam\relax}%
 \supsubtoks{}\def\@bi@lap{}\count@\@ne \bi@gint}
\slugcomment{submitted to ApJ}
\begin{document}
\title{Contamination of the Central Sunyaev--Zel'dovich Decrements in AMiBA Galaxy Cluster Observations}
\author{
Guo-Chin Liu\altaffilmark{1,2},
Mark Birkinshaw\altaffilmark{3},
Jiun-Huei Proty Wu\altaffilmark{5},
Paul T. P. Ho\altaffilmark{1,4}
Chih-Wei Locutus Huang\altaffilmark{5},
Yu-Wei Liao\altaffilmark{5},
Kai-Yang Lin\altaffilmark{1,5},
Sandor M. Molnar\altaffilmark{1}, Hiroaki Nishioka\altaffilmark{1},
Patrick M. Koch\altaffilmark{1},
Keiichi Umetsu\altaffilmark{1}, Fu-Cheng Wang\altaffilmark{5},
Pablo Altamirano\altaffilmark{1},
Chia-Hao Chang\altaffilmark{1},
Shu-Hao Chang\altaffilmark{1},
Su-Wei Chang\altaffilmark{1},
Ming-Tang Chen\altaffilmark{1},
Chih-Chiang Han\altaffilmark{1},
Yau-De Huang\altaffilmark{1},
Yuh-Jing Hwang\altaffilmark{1},
Homin Jiang\altaffilmark{1},
Michael Kesteven\altaffilmark{6},
Derek Kubo\altaffilmark{1},
Chao-Te Li\altaffilmark{1},
Pierre Martin-Cocher\altaffilmark{1},
Peter Oshiro\altaffilmark{1},
Philippe Raffin\altaffilmark{1},
Tashun Wei\altaffilmark{1},
Warwick Wilson\altaffilmark{6}
}
\altaffiltext{1}{Academia Sinica, Institute of Astronomy and Astrophysics,
P.O.Box 23-141, Taipei 106, Taiwan}
\altaffiltext{2}{Department of Physics, Tamkang University, 251-37
  Tamsui, Taipei County, Taiwan}
\altaffiltext{3}{Department of Physics, University of Bristol, Tyndall Ave, Bristol, BS8 1TL, UK}
\altaffiltext{4}{Harvard-Smithsonian Center for Astrophysics, 60 Garden Street, Cambridge,
MA 02138, USA}
\altaffiltext{5}{Department of Physics, Institute of Astrophysics, \& Center
for Theoretical Sciences, National Taiwan University, Taipei 10617, Taiwan}
\altaffiltext{6}{Australia Telescope National Facility, P.O.Box 76, Epping NSW
1710, Australia}

\begin{abstract}
We investigate the contamination of the Sunyaev--Zel'dovich (SZ) effect
for six galaxy clusters, A1689, A1995, A2142, A2163, A2261, and
A2390, observed by the Y. T. Lee Array for Microwave Background
Anisotropy during 2007. With the range of baselines used, we find 
that the largest effect (of order $13\%-50\%$ of the central SZ flux density)
comes from primary anisotropies in the cosmic microwave background
and exceeds the thermal noise in all six cases.
Contamination from discrete radio sources is estimated to be at a
level of $(3\%-60\%)$ of the central SZ flux density. 
We use the statistics of these contaminating sources to estimate and correct
the errors in the measured SZ effects of these
clusters. 
\end{abstract}

\section{Introduction}
Measurements of the Sunyaev-Zel'dovich effect \cite[SZE]{SZE} are
contaminated by foregrounds and backgrounds at a level that depends on
the angular scale of interest and the frequency of
observation. Primary anisotropies in the cosmic microwave background 
(CMB) from redshift $z \simeq 1100$ dominate on 
scales larger than a few arcminutes ($\ell \lsim 2000$) but have a
spectrum that differs significantly from the SZE, and so can be
removed by multiwavelength observations. Galactic 
contamination arises from synchrotron, free--free, and dust emission
and can also affect SZE measurements. Synchrotron and free--free emission
usually dominate at low radio frequencies, while dust emission is most
important at millimeter and submillimeter wavelengths, with a
crossover frequency of 
$60 - 70$~GHz. Emission from 
discrete radio sources is less important at high radio frequencies,
but can be problematic even at 90~GHz. Spatial filtering, by observing
at a finer angular resolution than the scale of the SZE, can mitigate
this problem. Finally, SZE data are commonly affected by the pickup of
signals from the ground, the atmosphere, and other local emitters,
which have to be carefully corrected for or avoided.

Array for Microwave Background Anisotropy (AMiBA; Ho et al.~2009; Chen
et al.~2009; Koch et al.~2009a; Wu et 
al.~2009; Li et al.~2010) will be subject to all these issues. AMiBA
was designed to 
operate from $86$ to $102$~GHz and to be sensitive to multipoles 
$800 \lsim\ \ell \lsim\ 2600$ with the initial close-packed
configuration of seven 0.6m antennas (see Figure 2 of Ho et al. 2009),
and to $ 1600 \lsim \ell \lsim\ 9000$ with the current upgraded configuration of
thirteen 1.2m antennas (Molnar et al. 2010).

 The observing frequency was chosen to
minimize contamination from non-SZE  while working within a
window of good atmospheric transmission. It is generally assumed that
radio source contamination is less of a problem for AMiBA than for
other instruments that operate at lower frequencies. However, sources with
inverted spectra may still cause difficulties, so it is
important to consider the radio environment for any AMiBA field. The
level of CMB contamination may be anomalously high in any
particular field, and this should also be investigated. The purpose
of this paper is to investigate, and correct for, these contaminations
in the SZE signals measured from the first-year AMiBA observation,
so as to provide unbiased SZE estimations for further science investigations
in our companion papers.

We do not, here, consider the effects of ground pick-up and other 
technical issues to do with the observing technique --- the two-patch
method used to deal with such problems is discussed elsewhere (Lin et
al.~2009, Wu et al.~2009). Rather we investigate the CMB and Galactic
emission environment (using data from the {\it Wilkinson Microwave
Anisotropy Probe}, {\it WMAP}) and the known centimeter-wave radio
source population 
(using source lists from several surveys) to estimate their effect
on SZE results for the six clusters of galaxies observed by AMiBA. We
describe our SZE analysis in Section~2, and discuss the errors
in the presence of the sources of contamination in Section~3. A summary
of the final results is given in Section~4.

\section{SZE Flux Analysis}

Between 2007 April and August AMiBA observed six galaxy clusters,
A1689, A1995, A2142, A2163, A2261 and A2390. These SZE 
data have been combined with X-ray information in the literature to
estimate the Hubble constant (Koch et al.~2009b) and study the
scaling relation (Huang et al.~2009, Liao et al.~2010).
In combination with Subaru weak lensing data they have also been used 
to study the gas fraction (Umetsu et al. 2009). Crucial to this
work is the proper estimation of the errors on the measured SZEs in
the presence of contaminating signals. In this section we describe
the basic analysis used to measure the key parameters of the cluster
SZEs. 

The fundamental observable for AMiBA is the visibility, which
is the Fourier transform of the sky intensity multiplied by the
primary beam or aperture function: 

\begin{equation}
{\cal V}(u, v) = g \int dx dy A(x,y) \Delta I(x,y)
                  e^{- 2 \pi i (u x + v y) } ,
\label{eq:visibility}
\end{equation}
where $g$ is a gain factor, which can be measured by calibration,
$(u,v)$ are components of the spatial separation vector of two
antennas in the array in units of the observing wavelength, 
$\lambda$, $A(x,y)$ is the primary beam pattern, 
$(x,y)$ are two components of angular position on the sky, measured
relative to the phase center (taken here to coincide with the
pointing center), and $\Delta I(x,y)$ is the distribution of surface
brightness of the sky. Here we have approximated the sky as being
flat. Details of the conversion from correlator data to visibilities
are described 
in Wu et al.\ (2009).

Given the SZE flux density in the sky, one can calculate the
visibility from Equation~(\ref{eq:visibility}). The SZE flux density
of a cluster depends on observing frequency, 
the physics of the cluster atmosphere (gas temperature and density
distributions, peculiar velocity, etc), and the angular diameter
distance, $D_A$. If we adopt a spherical isothermal $\beta$-model
\cite{betamodel}, the visibilities measured by an interferometer are
\begin{eqnarray}
  {\cal V}_{\rm SZE}(u, v) &= I_0 \int dx \, dy \, A(x,y)
                    \left( 1 + \frac{\theta^2}{\theta_0^2}
                    \right)^{-\frac{3}{2}\beta+\frac{1}{2}} \nonumber \\
                 & \times e^{- 2 \pi i (u x + v y) } \quad ,
  \label{eq:isze2}
\end{eqnarray}
where $I_0$ is the central surface brightness of the SZE, 
 $\theta^2 = (x - x_0)^2 + (y - y_0)^2$,
$(x_0,y_0)$ gives the sky position of the center of the cluster, 
$\theta_0$ is the angular core radius of the cluster, and $\beta$ is
the shape parameter.

For much of the SZE science \cite{birkinshaw} that we can currently
tackle with AMiBA we need to estimate the quantity $I_0$ by fitting
the model in Equation~(\ref{eq:isze2}) to the measured
visibilities. 
AMiBA visibilities are estimated in a process that includes
conversion from the wide-band, 4-lag, correlated analog signal
into visibilities in two frequency channels~(Wu et
al.~2009) after tests on the quality of the data as discussed by 
Nishioka et al.~(2009). In the hexagonal close-packed configuration
used for observing the six clusters, the seven elements
instantaneously yield 21 baseline vectors of three different
lengths. Eight platform rotation angles relative to the
pointing center are used to improve the $u-v$ sampling.

\begin{figure}
 \plotone{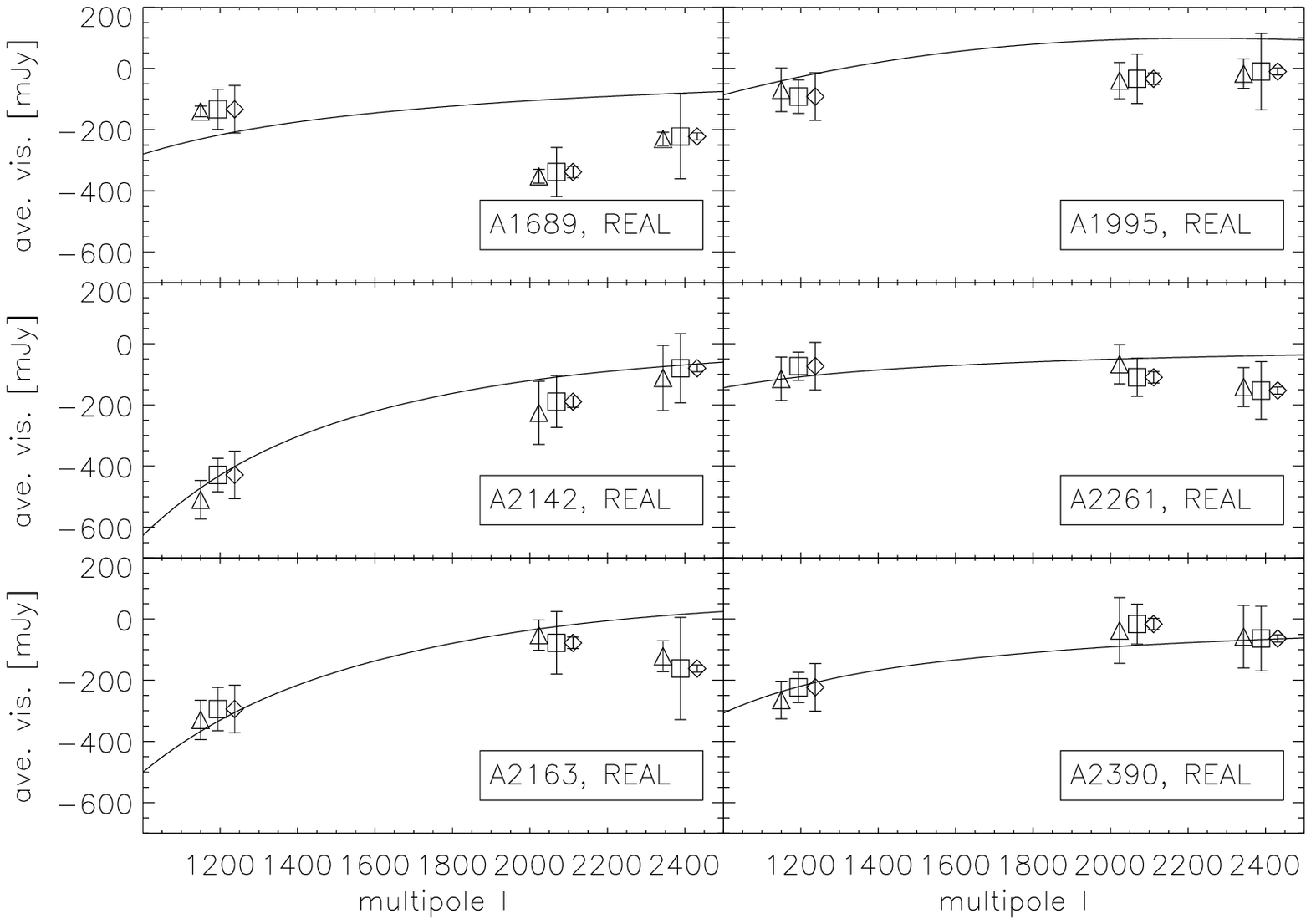}
 \plotone{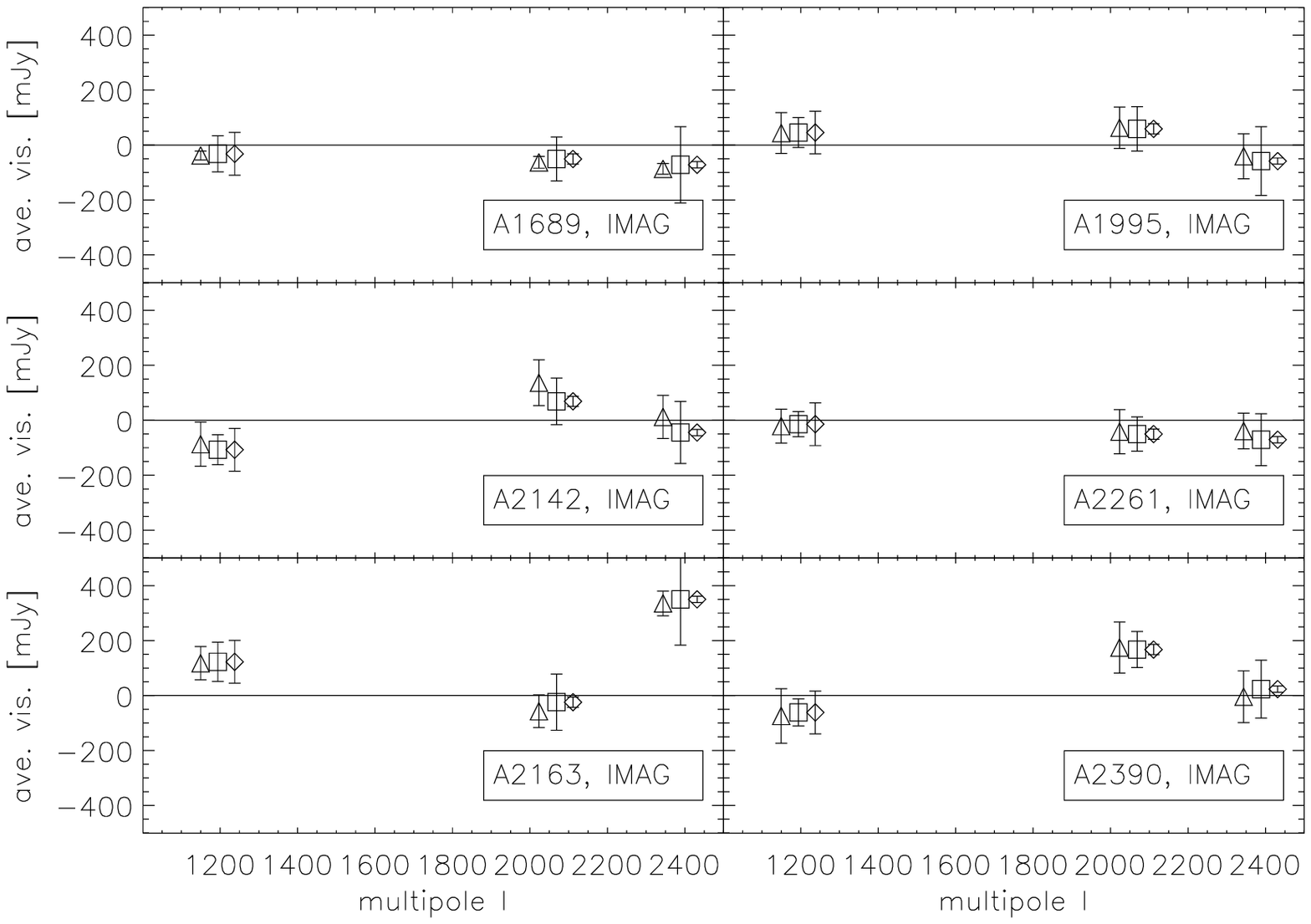}
 \caption{Averaged visibilities and best-fitting $\beta$-model 
  profiles for the six AMiBA galaxy clusters. The data points show
  error bars based on the 1-$\sigma$ instrumental noise (square), CMB
  (triangle) and radio sources (diamond). Data at same multipole are
  separated horizontally for better visualization.
}
 \label{fig:ave_vis}
\end{figure}

In Figure~\ref{fig:ave_vis}, we show the observed real and
imaginary parts of the visibilities as a function of
multipole $\ell=2\pi (u^2+v^2)^{1/2}$. The averaging in
azimuth uses $\sigma^{-2}$ noise weighting, and 
for each baseline length the information from 
the two frequency channels is combined.

We fit the visibility model in Equation~(\ref{eq:isze2}) by
minimizing $\chi^2$ with respect to $I_0$ and the location
$(x_0,y_0)$ of the cluster center. Because of the limited coverage in
$\ell$, the values of $\beta$ and $\theta_0$ can not be usefully
constrained by our data so we use values from the literature, which are
consistently used by our companion papers (Huang et al.~2009, Koch et
al.~2009b, Wu et al.~2009). 

The fitted results for $I_0$, $x_0$, and $y_0$ for the clusters are
given in Table~\ref{tbl:fit}. The errors were calculated from the
Hessian matrix in the usual manner. The best-fit visibility profiles
are shown with the data in Figure~\ref{fig:ave_vis}. 
At this stage no contribution to the errors from CMB or radio source
confusion has been taken into account. These effects are
the subject of the remainder of this paper. 

\begin{deluxetable}{lccc}
 \tabletypesize{\scriptsize}
 \tablecolumns{5}
 \tablecaption{
 \label{tbl:fit}
  Best-fit Values for $I_0$ and Offset of the Cluster centres, with
  1-$\sigma$ Uncertainties Ignoring the Effects of Contamination.}
 \tablewidth{0pt}
 \tablehead{\colhead{Cluster} & \colhead{$I_0$} 
                              & \colhead{Offset in R.A.} 
                              & \colhead{Offset in Decl.}\\
            \colhead{}        & \colhead{($\times 10^5$Jy ${\rm sr}^{-1}$)} 
                              & \colhead{(arcmin)} 
                              & \colhead{(arcmin)}
 }
 \startdata
  A1689 & -2.24 $\pm$ 0.44 &           -0.05$\pm$ 0.63 
                           & \phantom{-}0.02$\pm$ 0.63 \\
  A1995 & -3.38 $\pm$ 0.63 & \phantom{-}4.41$\pm$ 0.76 
                           & \phantom{-}0.38$\pm$ 0.68 \\
  A2142 & -2.16 $\pm$ 0.19 &           -0.64$\pm$ 0.42 
                           & \phantom{-}0.19$\pm$ 0.41 \\
  A2163 & -3.56 $\pm$ 0.37 & \phantom{-}2.53$\pm$ 0.47 
                           &           -2.36$\pm$ 0.53 \\
  A2261 & -1.46 $\pm$ 0.40 & \phantom{-}0.43$\pm$ 1.21 
                           & \phantom{-}1.31$\pm$ 1.19 \\
  A2390 & -2.42 $\pm$ 0.36 & \phantom{-}0.60$\pm$ 0.81 
                           &  -1.74$\pm$ 0.56
 \enddata
\end{deluxetable}

\section{Error Analysis}
\subsection{Large-scale CMB and Galactic-emission
 contamination}\label{sec:largecont} 

{\it WMAP} has recently released full-sky temperature and polarization
maps based on their five-year accumulated data. The angular
resolution of {\it WMAP} varies from $0^\circ.88$ to $0^\circ.22$ over five
bands from 23~to 94~GHz, and since the frequency range extends
up to the AMiBA operating band, the {\it WMAP} data provide a good check
on the level of CMB and Galactic contamination of the AMiBA data,
although at lower angular resolution.

We use the CMB map produced by the internal linear combination (ILC)
method that is independent of both external data and assumptions
about foreground emission, and estimate the Galactic foreground
using the maps of synchrotron, free--free, and dust foregrounds 
produced by the maximum entropy method~\cite{fg_wmap}. All four maps
are smoothed to $1^\circ$ scales, so that the principal indication
that can be obtained is only whether there are unusually ``hot'' or
``cold'' spots near the clusters or trailing fields, and so whether
there is likely to be additional noise power on the angular scales
to which AMiBA is sensitive.

We find no strong Galactic contamination in any of the six clusters,
with the total Galactic emission being typically an order of
magnitude fainter than the CMB signal (Table~\ref{tbl:mapcheck}). On the 
smaller angular scales  
to which AMiBA is sensitive, the Galactic emission will be even less 
important, since its power spectrum falls roughly as a power law, 
$C_\ell \propto \ell^{-\gamma}$, with $\gamma = 2.0 - 3.0$ 
\cite{tegmark}. 
Here the power spectrum is defined as
\begin{equation}
C_\ell = \frac{1}{2 \ell +1} \sum_m <a_{\ell m}^{*}
a_{\ell m}>
\end{equation}
with $a_{\ell m}$ being the expansion coefficient of the temperature field on
the spherical harmonics. Thus, we expect that Galactic contamination
will be two orders of magnitude fainter than the primordial CMB signal
as seen in the {\it WMAP} data.

We summarize the CMB surface brightness fluctuations in the
clusters and trailing fields on the angular scale of the clusters,
based on the {\it WMAP} ILC image, in Table~\ref{tbl:mapcheck}. 
The values range up to $\sim 25$~kJy ${\rm sr}^{-1}$ 
($\sim 100 \ \mu$K in temperature units). The level of 
primordial contamination at the degree scale in our data is suppressed
by the 45 arcmin separated two-patch subtraction technique that we
use, but nevertheless we might expect elevated levels of CMB noise in
fields with large degree-scale signals. This is consistent with the
observed deviations from the expected visibility curves
for the most contaminated clusters, A1689 and A2390.

\begin{deluxetable}{lcc}
 \tabletypesize{\scriptsize}
 \tablecolumns{3}
 \tablecaption{
  \label{tbl:mapcheck}
 Large-scale CMB and Galactic Emission ( Smoothed to $1^\circ$ Scales).}
 \tablewidth{0pt}
 \tablehead{\colhead{}&
   \colhead{CMB(lead;trail)}&
   \colhead{Galactic(lead;trail)}\\
  \colhead{Cluster}&
   \colhead{($\times 10^3$Jy ${\rm sr}^{-1}$)}&
   \colhead{($\times 10^3$Jy ${\rm sr}^{-1}$)}
 }
 \startdata
  A1689 & $          +24.1 ;           +22.3$ & $0.0 ; 0.0$ \\
  A1995 & $-\phantom{1}5.6 ;           +14.0$ & $0.0 ; 0.0$ \\
  A2142 & $+\phantom{1}8.5 ;           +14.8$ & $0.0 ; 0.0$ \\
  A2163 & $          +17.2 ; -\phantom{1}4.3$ & $1.1 ; 7.3$ \\
  A2261 & $-\phantom{1}0.2 ; +\phantom{1}4.2$ & $0.5 ; 0.4$ \\
  A2390 & $          -25.2 ;           -20.7$ & $4.0 ; 1.4$ 
 \enddata
\end{deluxetable}

\subsection{Small-scale CMB and Galactic-emission
 estimation}\label{sec:smallcont} 

Given the statistical properties of the CMB and Galactic emission, we
are able to estimate the rms fluctuations on each baseline. Because of
the two-patch observing strategy, we rewrite
Equation~(\ref{eq:visibility}) as 
\begin{eqnarray}
 {\cal V}(u_j, v_j, x_p, y_p) &=&  \int dx \ dy \ A(x-x_p,y-y_p) \Delta
 I(x,y) \nonumber\\
 &\ & \times e^{- 2 \pi i (u_j (x-x_p) + v_j (y-y_p))} \nonumber \\
 &=& \int du dv \tilde{A}(u_j -u, v_j -v) a(u, v)\nonumber \\
 &\ &\times e^{2 \pi i (ux_p+vy_p)},
 \label{eq:visibility2}
\end{eqnarray}
where $(x_p,y_p)$ is the pointing center of the corresponding field,
and $a(u,v)$ is the Fourier transform of surface brightness of the
sky. The two-point correlation function of $a({\bf{u}})$ is diagonal
because of the rotation invariant 
\begin{equation}
 <a({\bf{u}})a({\bf{w}})>=C_{\bf u} \, \delta^{(2)}({\bf{u}}-{\bf{w}}), 
\end{equation}
where $C_{\bf u}=(\partial B_\nu/\partial T)^2C_{\ell}$ and $\partial
B_\nu/\partial T$ converts from temperature to intensity. 
We have adopted the 
flat-sky approximation, which is valid for AMiBA, 
so that $\ell = 2\pi {\bf{u}}=2\pi \sqrt{u^2+v^2}$. 

The rms fluctuation in each baseline is then
\begin{eqnarray}
 <{\cal V}^2({u_j,v_j})>=\int \, du \, dv &\tilde{A}^2(u_j -u, v_j -v)
 \, C_{\ell}  \nonumber \\
  & \times \bigl( 1 - \cos [2\pi(u\Delta x+v\Delta y)] \bigr) \quad , 
 \label{rmspower}
\end{eqnarray}
where $\tilde{A}$ is the Fourier transform of the primary beam,
($\Delta x, \Delta y$) is the separation of the clusters and trailing
fields (about 45 arcmin in most of our data). Since the universe
is isotropic, we put $\Delta x=45$ and $\Delta y=0$ arcmin when we
calculate the rms fluctuation. The 1-$\sigma$ uncertainties caused by
the CMB at the $l$~ranges corresponding to AMiBA baselines are
plotted in Figure~\ref{fig:ave_vis} as shown by the diamond symbol.

\begin{figure}
 \plotone{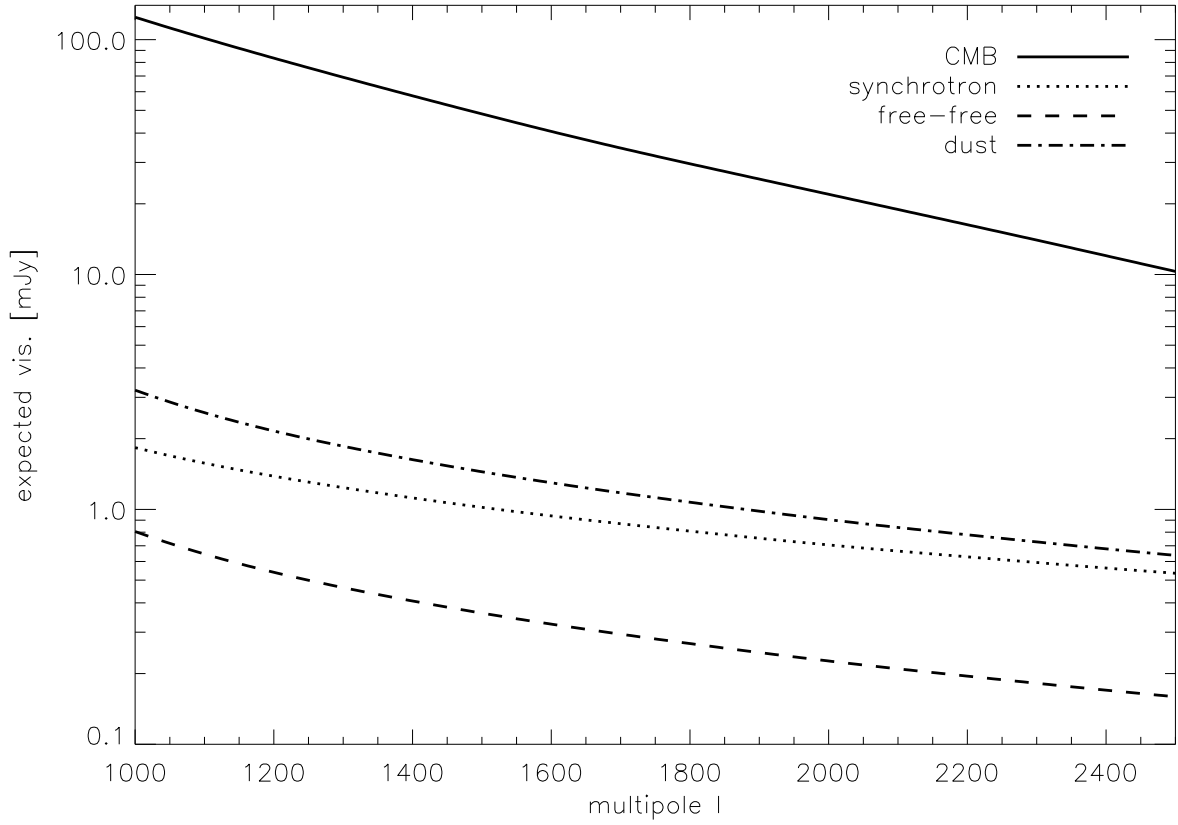}
 \caption{Expected response from CMB (solid line), synchrotron (dotted
  line), free--free (dashed line), and dust (dot-dashed) emission as a
  function of multipole $\ell$.
 }
 \label{fig:cmb_fg}
\end{figure}

For the Galactic emission, we use the middle-of-the-road (MID)
foreground model in Tegmark et al.~(2000), which is intended to be
realistic but conservative. We show the expected fluctuation from
three Galactic emission components in Figure~\ref{fig:cmb_fg}. The
Galactic emission lies about two orders of magnitude below the
contribution from the CMB. Other models in Tegmark et al.~(2000) change the
results by less than a factor of 2. These results are consistent with
what we estimated in Section~\ref{sec:largecont}. Therefore, we ignore the
contribution from Galactic emission in what follows. 

The CMB surface brightness fluctuations expected for a $\Lambda$CDM
model with cosmological parameters as estimated from the {\it WMAP} five-year
data alone (Dunkley et al.~2009) contribute
uncertainties of $80.6$, $20$, and $11.4$~mJy for the three averaged
AMiBA baselines and exceed the instrument noise for all 
baselines in all six clusters.

\begin{deluxetable}{lccc}
 \tabletypesize{\scriptsize}
 \tablecolumns{4}
 \tablecaption{
  \label{tbl:error}
  Errors and Systematic Offsets in the Estimates of $I_0$ Caused by
  CMB Fluctuations and Radio Sources. The final estimate
  for $I_0$ includes these contributions and supersedes the
  value in Table~\ref{tbl:fit}.
 }
 \tablewidth{0pt}
 \tablehead{
  \colhead{}&
   \colhead{$\sigma_{\rm CMB}$}&
   \colhead{Source Offset}&
   \colhead{Final $I_0$}\\
  \colhead{Cluster}&
   \colhead{($\times 10^5$Jy ${\rm sr}^{-1}$)}&
   \colhead{($\times 10^5$Jy ${\rm sr}^{-1}$)}&
   \colhead{($\times 10^5$Jy ${\rm sr}^{-1}$)}
}
\startdata
  A1689 &  0.53 & $-0.16 \pm 0.17$ & $-2.40 \pm 0.71$\\
  A1995 &  1.02 & $+0.28 \pm 0.27$ & $-3.10 \pm 1.22$\\
  A2142 &  0.30 & $-0.24 \pm 0.12$ & $-2.40 \pm 0.37$\\
  A2163 &  0.47 & $-0.01 \pm 0.13$ & $-3.16 \pm 0.61$\\
  A2261 &  0.77 & $-0.88 \pm 0.24$ & $-2.34 \pm 0.90$\\
  A2390 &  0.61 & $-0.51 \pm 0.31$ & $-2.93 \pm 0.77$
\enddata
\end{deluxetable}

To estimate the error in $I_0$ from CMB anisotropies, we repeat
the fitting of Section.~2 for 500~simulated CMB skies where the CMB
anisotropies have a power spectrum based on the 
cosmological  
parameters from the five-year {\it WMAP} data.  For each
simulation we subtract the simulated CMB visibilities from the data
and fit $I_0$, assuming that the central positions in
Table~\ref{tbl:fit} are unchanged. Our results for the noise
introduced in the values of $I_0$, $\sigma_{\rm CMB}$, are given in
Table~\ref{tbl:error}.

\subsection{Foreground Point Sources}

\begin{deluxetable}{lccccccc}
 \tabletypesize{\scriptsize}
 \tablecolumns{8}
 \tablecaption{
  \label{tbl:pts}
  Radio Sources in the Cluster Fields from $6$~cm Catalogs.
 }
 \tablewidth{0pt}
 \tablehead{
  \colhead{} &
    \colhead{R.A.} &
    \colhead{Decl.} &
    \colhead{$S_{6\ \rm cm}$}\tablenotemark{a}&
    \colhead{$S_{20\ \rm cm}$}\tablenotemark{b}&
    \colhead{$S_{3\ \rm mm}$}\tablenotemark{c}&
    \colhead{$\alpha$} &
    \colhead{$\Delta \theta$}\\
  \colhead{Cluster} &
    \colhead{(J2000)} &
    \colhead{(J2000)} &
    \colhead{(mJy)} &
    \colhead{(mJy)} &
    \colhead{(mJy)} &
    \colhead{}    &
    \colhead{(arcmin)} 
 }
 \startdata
  A1995&&&&&&&\\
   &14 52 50.8&57 46 58.0& 28& 25&36.0& 0.08&16.0\\
   &14 52 33.8&57 53 48.0& 21& 53& 2.3&-0.75& 9.3\\
   &14 52 16.9&58 13 28.0& 17& 53& 1.1&-0.91&11.5\\
   &14 57 41.2&57 57 03.5& 60&182& 4.3&-0.89&12.7\tablenotemark{*}\\
   &14 57 56.3&57 44 47.1& 56& 81&23.0&-0.3 &20.0\tablenotemark{*}\\
  A2142&&&&&&&\\
   &15 58 13.0&27 16 22.4& 39&107& 3.5&-0.81& 3.2\\
   &15 58 34.9&27 30 45.0& 21& ... &  ... &   ...&17.4\\
   &16 02 59.6&27 21 35.0& 83&364& 2.4&-1.19& 6.2\tablenotemark{*}\\
  A2163&&&&&&&\\
   &16 15 54.6&-6 08 44.0& 42& ...&  ...&   ...& 5.3\\
   &06 18 40.8&-6 17 18.0& 56& ...&  ...&   ...&10.0\tablenotemark{*}\\
  A2261&&&&&&&\\
   &17 22 24.2&32 01 24.3& 56&126& 8.1&-0.65& 6.3\\
   &17 26 19.4&32 19 19.5& 37&104& 3.2&-0.82&15.2\tablenotemark{*}\\
   &17 26 20.5&32 01 26.0& 29& ...&  ...&   ...&11.7\tablenotemark{*}\\
   &17 26 35.3&32 13 30.0&192&126& 522& 0.34&14.3\tablenotemark{*}\\
A2390&&&&&&&\\
   &21 54 40.9&17 27 53.0&294&294& 294& 0   &18.6\\
   &21 57 05.4&17 51 15.0& 81&270& 4.6&-0.97&10.5\tablenotemark{*}\\
   &21 56 43.8&17 22 49.0& 41&136& 2.3&-0.93&19.1\tablenotemark{*}
  \enddata
  \tablenotetext{a}{PMN and GB6~\cite{gregory, pmn}}
  \tablenotetext{b}{NVSS catalogs~\cite{nvss}}
  \tablenotetext{c}{Extrapolation to AMiBA frequency by power law}
  \tablenotetext{*}{Radio sources located in the trailing fields}
\end{deluxetable}

The emission from discrete sources will reduce the size of the 
SZE at 94~GHz if those sources are concentrated towards the target
cluster, but can cause an increase in the measured value of $I_0$ if
they are located in the trailing field (or lie in negative sidelobes 
of the synthesized beam). Since we expect to see more radio sources
toward the clusters, the measured SZE will normally be an 
underestimate of the true value \cite{coble}. 

Corrections for point source contamination can often be made by
combining arcminute-resolution SZE data with high-resolution data
(see, e.g., Lancaster et al.~2005; Udomprasert et al.~2004). However,
such supporting data are not available for AMiBA. 
Because the point source population at 90~GHz is also poorly known,
it is challenging to estimate the degree of contamination.
Thus, we take the following approach.

In assessing the contributions from known sources, we use the 
$4.85$ GHz GB6 and PMN catalogs \cite{gregory, pmn}. The GB6 survey
covered declinations $0^\circ < \delta < +75^\circ$ to a flux density
limit $S_{4.85} = 18$~mJy. The four declination bands in the PMN survey
from $-88^\circ < \delta < +10^\circ$ and have flux density
limits $S_{4.85} = 20 - 70$~mJy. We estimate the contamination for
fainter sources with significant low-frequency emission using 
the NVSS catalog at 1.4~GHz \cite{nvss}, which is complete for
$\delta > -40^\circ$ with flux density limit $2.5$~mJy. In each case,
we investigate the radio environments of the clusters by extracting
radio sources within $20^\prime$ of clusters and trailing
fields. The properties of these sources are summarized in
Table~\ref{tbl:pts}. We also 
estimate their flux densities at AMiBA's operating frequency
by simply assuming power law spectra of the form $S_\nu \propto
\nu^{\alpha}$ with $\alpha$ the spectral index, assuming that source
variability is not important.

In each field we find $20 - 30$ objects in the NVSS and from 0-3
objects in the GB6 or PMN lists. Most of these sources are too 
faint to affect the AMiBA data, but two GB6 sources are bright enough 
to cause problems. The first is a source with $S_{4.85} = 192$~mJy 
that lies
$14^\prime\llap{.}3$ from the center of the trail patch of A2261.
The other has $S_{4.85} = 294$~mJy and lies $18^\prime\llap{.}6$ 
from the center of A2390. 
However, the VLA calibrator catalog suggests 
that both sources have falling spectra above 8~GHz, with 
predicted 94~GHz flux densities of $130 \pm 20$ and $90 \pm 20$~mJy
if the high-frequency spectra are power laws and variability is not
important. Both sources lie beyond the half-power point of the 
AMiBA beam (at the $30 \%$ and $13 \%$ levels for offsets of
$14^\prime\llap{.}3$ and $18^\prime\llap{.}6$, respectively; Wu et
al.~2009). We take account of these sources by performing a
simultaneous fit for the cluster SZEs and 
the spectral indices of all sources with 4.85 GHz flux densities
in Table~\ref{tbl:pts} for these two clusters. We found 
spectral indices of $-0.02 \pm 0.13$ (implying a 3 mm flux density
$S = 203_{-67}^{+100}$~mJy) and $0.04 \pm 0.03$ (implying a 3 mm flux
density $S = 328_{-198}^{+500}$~mJy) for the problem sources in the
trail patch of A2261 and the main A2390 field, respectively. Although
the sources are barely detected, we expect them to significantly
affect the fitted value of $I_0$. 

None of the catalogued point sources is detected by AMiBA (at
an rms sensitivity of $50$~mJy) in the residual maps formed by
subtracting the best-fit cluster SZE models from the
visibility data. Therefore, we adopt a statistical method of
estimating the contamination from the known point sources. This
involves simultaneously estimating the spectral indices of
known sources from $4.85$ to $94$~GHz and the value of $I_0$ for the
cluster, as we did for A2261 and A2390. We do not use any prior on the
spectral indices because the results are sensitive to the assumed
distribution. 

We cannot apply this method for the NVSS sources because
the problem is generally underdetermined: there are too many NVSS
sources per field. Instead we
account for the contribution of NVSS sources using Monte Carlo 
simulations. In each of 500 runs, we pick a set of spectral indices
for the sources from the {\it WMAP} distribution, predict the 94~GHz
flux densities (forcing no source to exceed the GB6 or PMN limit),
and estimate the visibilities. We then subtract these simulated 
visibilities from the AMiBA data, and re-fit the value of $I_0$ and
the spectral indices of the sources in the 4.85 GHz catalogs. We
expect this to be an upper limit on the level of confusion 
from foreground and background radio sources not represented
in the source surveys to date. 

The effects of point sources on the value of $I_0$ are summarized in
Table~\ref{tbl:error}. We also show the effect of the offset 
induced by point sources in Figure~\ref{fig:ave_vis} in the values
marked by triangles. The value of $I_0$ increases by less than
its random error in the four clusters without strong 
4.85 GHz contaminating sources, but significant changes are seen for
A2261 and A2390. In these cases, follow-up observations of the sources
are highly desirable. The additional noise contributed by the
population of point sources is, however, always considerably less than
the noise from CMB anisotropies.

As pointed out by Giommi and Colafrancesco (2004), some unresolved
sources with flat or inverted spectra may contaminate our SZ signal
at 94~GHz. Further, these sources are mostly composed of the blazars. 
However, we conclude that the contribution of blazars to our
observation is negligible by the analytical method as follows.  

The visibility observed by point sources to each baseline is
\begin{equation}
{\cal V}(u, v)=\sum_i S_i (\nu) A(x_i,y_i) e^{- 2 \pi i (u x_i + v y_i) },
\end{equation}
where $ S_i (\nu)$ is the flux density of the $i$th radio source. Assuming
the radio sources are spatially distributed like a Poissonian random
sample, the averaged visibility 
for each
baseline is expected to be zero due to the  two-patch subtraction
technique. The rms fluctuation in each baseline then is
\begin{eqnarray}
<{\cal V}^2({u,v})>&=&\frac{1}{\Omega}<\sum_i S_i^2(\nu)>\int dx
\ dy \ A^2(x,y) \nonumber \\
&=& \int_{S_{\rm min}}^{S_{\rm max}} dS S^2\frac{dN(S)}{dS} \int d\alpha
\, p(\alpha|S,\nu_0) \, \left( \frac{\nu}{\nu_0} \right)^2\int dx
\, dy \, A^2(x,y)
\label{eq:rmspt}
\end{eqnarray}
where $\Omega$ is the normalized solid angle, $S$ is the source flux
density, and $dN/dS$ is the differential source count as a function of
flux. Typically, the differential source count is described by a power law,
i.e., $dN/dS=N_0 S^{-\gamma}$, where $N_0$ is a normalization
parameter and $\gamma$ is the power-law index. Since the differential
source count of radio sources is well known only at low frequency, we are
forced to rely on models to extend it to AMiBA's operating frequency. 
   
We use the Equation~(\ref{eq:rmspt}) to estimate the contribution from
blazars on each baseline, and increase the resulting noise by $\sqrt{2}$
to take account of our two-patch subtraction technique.
The source counts and the spectral energy distribution of blazars are well
studied by Giommi and Colafrancesco (2004) and Giommi et al.~(2006,
2009), and we adopt the values of $N_0$ and 
$\gamma$ in Giommi et al.~(2006) based on the combination of data from
several radio and other surveys. We take the sources contributing in
the AMiBA band to have the same distribution of spectral indices as those
given by Giommi and Colafrancesco (2004), based on data 
between $5$ and $150$~GHz for a sample of
135 blazars from the NVSS-RASS $1$~Jy survey (Giommi et al.~2002)
and the $0.1$ and $0.05$~Jy points of the Deep X-Ray Radio Blazar
Survey. We integrate Equation~(\ref{eq:rmspt}) from $S=0$ to
the flux density limits of GB6 ($18$~mJy) and PMN ($40$~mJy). We find 
that the rms contribution of blazars on each baseline is $13$~mJy for
A1689 and A2163 and $8.9$~mJy for the other four clusters. These
contributions are negligible compared with the SZE flux densities
found for our cluster sample (about 150-300~mJy).

\section{Discussion and Conclusion}

The levels of contamination estimated in this paper are 
summarized in Table~\ref{tbl:error}. Except for A2261, the dominant
source of confusion is the primordial anisotropy of the CMB, which
adds noise at the level of $(13\% - 50\%)$ of the value of $I_0$,
exceeding the AMiBA system noise in every case. Galactic foreground
emission is always negligible.

Known point sources affect the amplitudes of the SZEs of
the clusters by less than the combined system and CMB noises in each
case except A2261, 
but ignoring the sources would systematically bias the estimates of $I_0$
since there are more radio sources towards the clusters than in the
background fields. A2261 suffers particularly substantial
contamination: both from the strong sources in the
4.85 GHz catalog (which cause a change $\Delta I_0 \sim -0.35 \times
10^5$~Jy ${\rm sr}^{-1}$) and from faint sources in the NVSS catalog
(which cause a further effect $\Delta I_0 \sim -0.53 \times 10^5$~Jy
${\rm sr}^{-1}$).   

Several other workers have reported on the discrete source
environments of these clusters. Reese et al.~(2002) report on radio
sources at 28.5~and 30~GHz for A1689, A1995, A2163 and A2261. All of
these sources are in NVSS, but none is bright enough at 5~GHz to
appear in the GB6 or PMN catalogs: the brightest is a 
10~mJy source in A2261. The VSA source-subtractor also reports on
sources in A2142~\cite{katy}: all are included in the NVSS catalog, and 
the contamination in the AMiBA data that they generate is small (less
than 3\% of $I_0$).

Tucci et al.~(2008) have studied radio source spectra from 
1.4 to 33 GHz. They found that, in general, the spectra are not well
described by a single power law: the low-frequency spectra are
usually steeper than the high-frequency spectra. Spectral index
studies from 1.4 to 28.5~GHz~(e.g., Coble et al. 2007) and
23 to 94~GHz \cite{wmap_ps} support this behavior. Our study,
which extrapolated from 1.4~GHz (or 4.85~GHz) to 94~GHz using the
{\it WMAP} spectral index distribution may therefore overestimate the
contamination from radio sources. 

Our best estimates of $I_0$, taking contamination into account, are
shown in the last column of Table~\ref{tbl:error}, where we show
the total uncertainty including the contributions from
instrument noise, CMB, and point sources. The values given in this
table are those used in the science analyses of our companion papers.

We acknowledge the extensive use we have made of data from the
{\it WMAP} satellite, and thank Dr.~L.~Chiang for fruitful discussions on
Galactic emission and the {\it WMAP} data. We also appreciate discussions
with Drs.~N.~Aghanim and S.~Matsushita on radio source properties.
Support from the STFC for MB is acknowledged. This work is partially
supported by the National Science Council of Taiwan under grants 
NSC97-2112-M-032-007-MY3 and NSC97-2112-M-001-020-MY3.

\end{document}